\documentclass[11pt]{article}

\usepackage[margin=1in]{geometry}
\usepackage[T1]{fontenc}
\usepackage{lmodern}
\usepackage{amsmath,amssymb}
\usepackage{amsthm}
\usepackage{booktabs}
\usepackage{graphicx}
\usepackage[hidelinks]{hyperref}
\hypersetup{
  pdftitle={Algorithms for Collaborative Truck-Multi-Drone Delivery in the En-Route Model},
  pdfauthor={Jerry Mathew Oommen and Kelin Luo},
  pdfkeywords={truck-drone delivery, approximation algorithms,
               dynamic programming},
}

\AtBeginDocument{%
  \setlength{\abovedisplayskip}{5pt plus 2pt minus 2pt}%
  \setlength{\belowdisplayskip}{5pt plus 2pt minus 2pt}%
  \setlength{\abovedisplayshortskip}{2pt plus 1pt}%
  \setlength{\belowdisplayshortskip}{3pt plus 1pt}%
}
\makeatletter
\renewcommand\paragraph{\@startsection{paragraph}{4}{\z@}%
  {1ex \@plus1ex \@minus.2ex}%
  {-1em}%
  {\normalfont\normalsize\bfseries}}
\makeatother

\theoremstyle{plain}
\newtheorem{theorem}{Theorem}
\newtheorem{lemma}{Lemma}
\newtheorem{corollary}{Corollary}
\newtheorem{definition}{Definition}
\theoremstyle{remark}
\newtheorem{remark}{Remark}

\DeclareMathOperator{\ret}{ret}
\DeclareMathOperator{\es}{es}
\DeclareMathOperator{\ls}{ls}

\title{\vspace{-1.2em}Algorithms for Collaborative Truck-Multi-Drone Delivery in the En-Route Model}
\author{Jerry Mathew Oommen\\[0.35em]
        {\small University at Buffalo}\\[0.15em]
        {\small E-mail: \href{mailto:jerryoommen07@gmail.com}{\texttt{jerryoommen07@gmail.com}}}
        \and
        Kelin Luo\\[0.35em]
        {\small University at Buffalo}\\[0.15em]
        {\small E-mail: \href{mailto:kelinluo@buffalo.edu}{\texttt{kelinluo@buffalo.edu}}}}
\date{\vspace{-1.0em}}

\begin{document}
\maketitle
\vspace{-1.5em}

\begin{abstract}

We study collaborative truck--multi-drone delivery in the en-route model, in which a truck moves continuously along a fixed straight-line path while drones serve \(n\) off-route deliveries.
For maximum coverage with \(k\) drones, we prove that Sequential Greedy achieves an approximation guarantee of \(1-(k/(k+1))^k\), which approaches \(1-e^{-1}\).
For \(k=2\), our analysis is tight, with a matching \(5/9\) instance.
We refute the claimed exchange property for proper instances.
For a new class of swap-monotone instances with distinct \(x\)-coordinates, we give an exact \(O(n^2)\)-time dynamic program for single-drone maximum coverage.
For minimum fleet size, we prove NP-hardness and give an \(O(1+\log(n/k^\ast))\)-approximation using an integer residual bound, where \(k^\ast\) is the optimum fleet size.
\end{abstract}

\vspace{0.3em}

\textbf{Keywords:} truck-drone delivery;  
approximation algorithms; dynamic programming

\section{Introduction}
\label{sec:intro}
A truck paired with one or more drones can cut both the cost and the time of
last-mile delivery: the truck carries the payload along a route while drones leave
the truck to serve individual deliveries and fly back. 
The main challenge is timing: each drone must launch from and rendezvous with a
\emph{moving} truck while remaining within its flight range. Most truck-drone research optimizes the truck's \emph{route}: Murray and Chu~\cite{Murray2015} introduced the Flying Sidekick TSP, and subsequent work considers multi-drone fleets~\cite{Murray2020,Boysen2021} and exact dynamic programs~\cite{Agatz2018,Bouman2018}. In contrast, we take a complementary view by fixing the truck route and asking \emph{which drone flies when}. This is the \emph{En Route} model of Krizanc et al.~\cite{Krizanc2024EnRoute}, in which the truck moves steadily along an infinite line and each delivery induces an interval of feasible launch positions from which a drone can serve the delivery and return to the truck in time. This interval is determined by a reachability ellipse.
Krizanc et al.~\cite{Krizanc2024EnRoute} study the single-drone case, prove that it is strongly NP-hard, and provide approximation algorithms.
We extend the model to a fleet of multiple drones with disjoint delivery assignments, and consider two problems.
\begin{itemize}\setlength{\itemsep}{1pt}
  \item \textbf{Maximum coverage} (fixed fleet): given \(k\)
        drones, serve as many of the \(n\) deliveries as possible.
  \item \textbf{Minimum fleet size} (full coverage): find the minimum number
        \(k^\ast\) of drones needed to serve every delivery.
\end{itemize}
The first problem fixes the available resources and maximizes service, whereas the second fixes the service requirement and minimizes resources. These are complementary variants of the same model. 

In this paper, we make the following contributions. First, we extend the existing truck-and-single-drone maximum coverage model~\cite{Krizanc2024EnRoute} to the multi-drone setting, where the objective is to maximize the number of deliveries served by $k$ drones. For Sequential Greedy, we sharpen the standard approximation guarantee by charging missed deliveries separately for each drone in an optimal solution. This yields the bound
$
1-\left(1-\frac{1}{k+1}\right)^k
$. Using the known single-drone algorithm as a black box gives a limiting
guarantee of \(1-e^{-1/2}\). Our direct analysis improves this to
\(1-e^{-1}\). For \(k=2\), our analysis is tight, with a matching \(5/9\) instance.

Second, we revisit a restricted class of instances, called proper instances, that has been studied in the existing literature~\cite{Krizanc2024EnRoute}. We give counterexamples to the adjacent-swap claim for proper instances~\cite[Lem.~5]{Krizanc2024EnRoute}. Motivated by this gap, we define the class of swap-monotone instances, for which adjacent out-of-order deliveries can be swapped without losing feasibility or delaying completion. Our prefix recurrence reduces the time for optimizing over \(x\)-ordered schedules from \(O(n^3)\)~\cite[Thm.~4]{Krizanc2024EnRoute} to \(O(n^2)\). It gives the exact single-drone optimum on swap-monotone instances with distinct \(x\)-coordinates.

Finally, we study the minimum-fleet problem for full coverage. We derive NP-hardness from the earlier single-drone reduction. For Sequential Greedy, our charging analysis replaces the black-box logarithmic coefficient \(2\) by \(1+1/k^\ast\) and yields a sharper integer residual bound. This gives an \(O(1+\log(n/k^\ast))\)-approximation, and the integer bound is tight for \(n=5\) and \(k^\ast=2\).

\section{Model and Preliminaries}
\label{sec:model}
The truck departs the origin at time $0$ and moves rightward along the positive
$x$-axis at unit speed, so its position at time $t$ is $(t, 0)$. A fleet of multiple drones, each with speed
$v>1$ and flight range $R$, serves a set of off-route deliveries
$D=\{d_1,\dots,d_n\}$. Delivery $d_i$ has location $(x_i,y_i)$, which may be
shared by other deliveries. A drone dispatched to serve a delivery launches from
the truck's current position, flies to the delivery location, and returns to
the moving truck.
Each sortie serves exactly one delivery; after returning, the drone may be
dispatched again. Recharge, pickup, and drop-off times are negligible (zero).
All drones start on the truck at time \(0\) and operate independently.
Simultaneous launches and returns are allowed.

\paragraph{Feasibility windows and return time.}
For a delivery \(d\) at \((x,y)\), the \emph{feasibility window} is the set of
truck positions from which a drone can be launched, serve \(d\), and return to
the truck within its flight range \(R\). We denote this interval by
$
W(d)=[\es(d),\ls(d)],
$ 
where \(\es(d)\) is the \emph{earliest starting time} and \(\ls(d)\) is the
\emph{latest starting time}. Following~\cite{Krizanc2024EnRoute}, these values are
\[
\es(d)=x-\frac{R}{2v}-x',
 \ 
\ls(d)=x-\frac{R}{2v}+x',
\]
where
$
x'=\frac{R}{2}\sqrt{1-\frac{y^2}{m_v^2}},
 \
m_v=\frac{R}{2v}\sqrt{v^2-1}.
$
Here \(m_v\) is the minor radius of the reachability ellipse.  If \(y=m_v\), then \(x'=0\), and the feasibility window collapses to the
single time
$
\es(d)=\ls(d)=x-\frac{R}{2v}.
$
We assume every delivery is individually serviceable after time \(0\):
\(0<y\le m_v\) and \(\ls(d)\ge0\). Unserviceable deliveries may be deleted
from a maximum-coverage instance; in a minimum-fleet instance they make full
coverage infeasible. 
For a launch time \(s\in W(d)\), the
\emph{return time} \(\ret(s,d)\) is the earliest time at which the drone can rejoin the
moving truck after serving \(d\), i.e., 
\[
\ret(s,d)=
s+
\frac{
s+av-x+
\sqrt{b^{2}-s(v^{2}-1)(b+s+av-x)}
}{
v^{2}-1
},
\]
where
$
a=\sqrt{y^{2}+(s-x)^{2}}, \
b=sv^{2}+av-x.
$
This expression is given in~\cite[Lem.\,1]{Krizanc2024EnRoute}. A dispatch of
\(d\) starting at time \(s\) is feasible if and only if \(s\ge0\) and
\(s\in W(d)\).

\section{Problem I: Maximum Coverage with a Fixed Fleet}
\label{sec:coverage}
In this problem, the number \(k\) of drones is fixed, and the objective is to
maximize the number of deliveries served. Each delivery may be served at most
once, and the schedule assigned to each drone must be feasible.

Krizanc et al.~\cite{Krizanc2024EnRoute} establish strong NP-hardness for the
single-drone maximum-coverage problem. This immediately yields the following
corollary.

\begin{corollary}\label{cor:nphard-cov}
Maximum coverage on general instances is strongly NP-hard even for the fixed
fleet size \(k=1\), and therefore remains strongly NP-hard when \(k\) is part
of the input.
\end{corollary}

\paragraph{Sequential Greedy.}
We use the single-drone algorithm of~\cite{Krizanc2024EnRoute} as a subroutine.
This algorithm is an \(\tfrac12\)-approximation. Starting
from time \(s=0\), it repeatedly selects, among unserved deliveries satisfying
\(\es(d)\le s\le \ls(d)\), the one with the smallest return time \(\ret(s,d)\).
It then advances \(s\) to this return time and drops expired deliveries from the
current pass. If no unserved delivery is currently launchable, the algorithm
advances \(s\) to the smallest value of \(\es(d)\) among the remaining deliveries.

Our Sequential Greedy algorithm applies this single-drone subroutine
once for each drone. After each iteration, the deliveries assigned to the current drone are removed, so
the resulting schedules have disjoint delivery assignments.
A standard maximum-coverage argument~\cite{Nemhauser1978Submodular} shows that each round closes at least an \(\alpha/k\) fraction of the remaining gap. Iterating over \(k\) rounds yields the following guarantee.

\begin{theorem}
\label{thm:seq-greedy-alpha}
If the single-drone subroutine is an \(\alpha\)-approximation, then Sequential
Greedy is a
$
1-\left(1-\frac{\alpha}{k}\right)^k
$-approximation algorithm.
\end{theorem}
\begin{remark}
Using the \(\alpha=\tfrac12\) approximation of~\cite{Krizanc2024EnRoute}, the
guarantee becomes
$
1-\left(1-\frac{1}{2k}\right)^k,
$
which converges to \(1-e^{-1/2}\approx 0.393\) as \(k\to\infty\).
\end{remark}

Each greedy pass closes at least a \(1/(k+1)\) fraction of the remaining gap
to the \(k\)-drone optimum. Lemma~\ref{lem:missed-injection} directly matches
each missed delivery to the first greedy return strictly after its reference
launch and proves that this return lies within its reference flight.
Nonoverlapping flights on each optimal drone allow at most \(k\) charges per
return. Lemma~\ref{lem:firstpass} aggregates these charges and counts shared
deliveries only once to obtain the multi-drone improvement.

\begin{lemma}[Missed-delivery injection]\label{lem:missed-injection}
Let \(A\) be any feasible single-drone schedule, and let \(G\) be the schedule
produced by one application of earliest-return Greedy to a delivery set
containing every delivery served by \(A\). Identifying each schedule with its
set of served deliveries, \(A\setminus G\) injects into \(G\).
\end{lemma}
\begin{proof}
Consider a delivery \(p\in A\setminus G\) whose reference flight launches at
\(s\) and returns at \(r\). Let \(g\) be the greedy flight whose return \(r'\)
is the first one strictly after \(s\), and let \(u\) be its launch. This flight
exists and \(u\le s\) because \(p\) is launchable at \(s\), so Greedy cannot be
idle then. If a greedy flight returns exactly at \(s\), its next flight
launches at \(s\).

Suppose \(r'>r\), and let \(z\) be the location of \(p\). The triangle inequality
and the reference flight give
\[
 \lVert z-(u,0)\rVert+\lVert z-(r,0)\rVert
 \le (s-u)+v(r-s)\le v(r-u).
\]
Thus a flight to \(p\) launched at \(u\) returns by \(r\). Its duration is less
than \(r'-u\le R/v\), so it is feasible and returns strictly before \(g\).
Since \(p\) is still unserved, this contradicts Greedy's earliest-return
choice. Therefore \(s<r'\le r\).

Map \(p\) to the greedy delivery returning at \(r'\). The reference flight
intervals \((s,r]\) are pairwise disjoint, so two missed deliveries cannot
map to the same greedy return. This proves the injection.
\end{proof}

Let \(S_i\) denote the set of deliveries assigned by Sequential Greedy to drone
\(i\). For \(P\subseteq D\), let \(\mathrm{OPT}(k\text{ on }P)\) denote the
maximum coverage achievable by \(k\) drones, optimized afresh on \(P\).

\begin{lemma}[Greedy-schedule bound]\label{lem:firstpass}
For any delivery set \(P\subseteq D\), let \(S\) be the set served by one
greedy pass on \(P\), and let \(m=\mathrm{OPT}(k\text{ on }P)\). Then
\(m\le(k+1)|S|\).
\end{lemma}
\begin{proof}
Let \(O_1,\dots,O_k\) be optimal disjoint schedules, allowing empty schedules
if the optimum uses fewer than \(k\) drones. Each is a feasible reference, so
Lemma~\ref{lem:missed-injection} gives \(|O_j\setminus S|\le|S|\) for every
\(j\), and hence \(\sum_j|O_j\setminus S|\le k|S|\). Moreover, the \(O_j\)
are disjoint, so the sets \(O_j\cap S\) are disjoint subsets of \(S\) and
\(\sum_j|O_j\cap S|\le|S|\). Adding the counts of deliveries already served
by \(S\) and those missed by \(S\) gives
\[
 m=\underbrace{\sum_j|O_j\cap S|}_{\le |S|}
   +\underbrace{\sum_j|O_j\setminus S|}_{\le k|S|}
 \le(k+1)|S|.
\]
The shared \(|S|\) term is the gain over applying the single-drone
factor-two bound separately to all \(k\) reference schedules.
\end{proof}

\begin{theorem}[Sequential Greedy guarantee]
\label{thm:seq-greedy}
For any delivery set \(D\), Sequential Greedy satisfies
\[
\frac{\lvert T_k\rvert}{\mathrm{OPT}(k\text{ on }D)}
\ge
1-\left(\frac{k}{k+1}\right)^k,
\]
where \(T_k=\bigcup_{i\in[k]} S_i\).
Consequently, the approximation ratio approaches
\(1-e^{-1}\approx 0.632\) as \(k\to\infty\).
\end{theorem}

\begin{proof}
Let \(O_1,\ldots,O_k\) be the schedules of an optimal solution on \(D\), and
put \(O=\bigcup_{j=1}^k O_j\) and \(m=|O|\). Define
$
T_t=\bigcup_{i=1}^{t}S_i$ and  
$
T_0=\emptyset.
$
The residual set is \(V_t=D\setminus T_t\), with \(V_0=D\).
Restricting the fixed schedules \(O_1,\ldots,O_k\) to \(V_t\) preserves
feasibility by retaining their launch times. These schedules cover
\(O\setminus T_t\), giving a lower bound on the residual optimum:
\[
\mathrm{OPT}(k\text{ on }V_t)
\ge
|O\setminus T_t|
\ge
m-|T_t|.
\]
Since \(S_{t+1}\) is the first greedy pass on \(V_t\),
Lemma~\ref{lem:firstpass} gives
\[
\mathrm{OPT}(k\text{ on }V_t)
\le
(k+1)|S_{t+1}|.
\]
Hence,
\[
|S_{t+1}|
\ge
\frac{m-|T_t|}{k+1}.
\]
Because \(S_{t+1}\subseteq V_t\), the sets \(S_{t+1}\) and \(T_t\) are
disjoint. Thus,
\[
|T_{t+1}|
=
|T_t|+|S_{t+1}|
\ge
|T_t|+
 \frac{m-|T_t|}{k+1}.
\]
Let \(\Delta_t=m-|T_t|\). 
\[ 
\Delta_{t+1}
\le 
\Delta_t- \frac{\Delta_t}{k+1}
= 
\frac{k}{k+1}\Delta_t.
\]
Iterating the above inequality for \(k\) passes and since $\Delta_0=m$, we get 
\[
|T_k|
=
m-\Delta_k
\ge m-
m\left(\frac{k}{k+1}\right)^k
=
m\left(1-\left(\frac{k}{k+1}\right)^k\right),
\]
which proves the claimed approximation guarantee.
\end{proof}

\paragraph{Tightness Discussion.}
Krizanc et al.~\cite{Krizanc2024EnRoute} show that the
\(\alpha=\tfrac{1}{2}\) approximation guarantee for single-drone Greedy is
tight. For \(k=2\), we similarly construct an instance showing that the bound
in Theorem~\ref{thm:seq-greedy} is tight.
Set \(R=5/2\) and \(v=10\), so that
\(m_v=\tfrac{R}{2v}\sqrt{v^2-1}=\sqrt{99}/8\), and take nine deliveries:
\[
\begin{array}{c|ccccccccc}
 &d_1&d_2&d_3&d_4&d_5&d_6&d_7&d_8&d_9\\\hline
x&.01&.05&.20&.22&.41&.44&.92&.93&.99\\
y&1.04&1.23&1.14&1.24&1.24&1.24&.46&.95&.68
\end{array}
\]
Substitution in the launch-window and return-time formulas gives the
Sequential Greedy decisions below.
\[
\begin{array}{c|c|c|c}
\text{schedule}&\text{delivery}&\text{launch}&\text{return}\\\hline
S_1&d_7&0&.189210358\\
 &d_9&.189210358&.385265953\\
 &d_8&.385265953&.595492734\\\hline
S_2&d_1&0&.209908702\\
 &d_3&.209908702&.440420608
\end{array}
\]
At each of these five launches the selected return is strictly smaller than
every other eligible return. All greedy comparisons are
strict, and no remaining delivery is feasible after either pass. Hence,
$ |S_1|+|S_2|=5.
$

In contrast, the following two feasible schedules serve all nine deliveries:
\[
\begin{array}{c|c|c|c}
\text{schedule}&\text{delivery}&\text{launch}&\text{return}\\\hline
O_1&d_2&0&.247679968\\
 &d_6&.247679968&.497294812\\
 &d_3&.497294812&.741306307\\
 &d_7&.741306307&.837391089\\
 &d_9&.837391089&.975098840\\\hline
O_2&d_4&0&.249972711\\
 &d_5&.249972711&.499322354\\
 &d_1&.499322354&.741402293\\
 &d_8&.741402293&.933256818
\end{array}
\]
This proves that the \(5/9\)-approximation guarantee for two-drone Sequential Greedy is tight. 

\subsection{Counterexamples and faster dynamic programming for restricted instances}
\label{sec:restricted-instances}
We next revisit the restricted instances considered in~\cite{Krizanc2024EnRoute}
and develop a faster dynamic program for a closely related class. First, we show
that a structural claim used in~\cite{Krizanc2024EnRoute} is not valid. We then
introduce swap-monotonicity and give an exact \(O(n^2)\)-time single-drone
algorithm for swap-monotone instances with distinct \(x\)-coordinates.

\paragraph{Counterexamples.}
Krizanc et al.~\cite[Def.~2]{Krizanc2024EnRoute} define the class of proper
instances. In such an instance, the location of \(d_j\) lies outside the launch
triangle of \(d_i\), with vertices \((\es(d_i),0)\), \((x_i,y_i)\), and
\((\ls(d_i)+R/v,0)\), for \(i\ne j\), and no feasibility window is contained
in another. They further claim that
swapping two consecutive deliveries served out of \(x\)-order preserves feasibility and does not
increase completion time~\cite[Lem.~5]{Krizanc2024EnRoute}. The proof
distinguishes two cases at the common free time: in Case~1, the smaller-\(x\)
delivery returns no later than the larger-\(x\) delivery, whereas Case~2 assumes
the reverse strict inequality. Appendix~\ref{app:case2-counterexamples} gives
two counterexamples for Case~2.

For Case~1, take \(R=0.4\), \(v=1.6\), and two deliveries at
\((0.25,0.02)\) and \((0.30,0.02)\). Equal altitude and distinct \(x\)-coordinates
make the instance automatically proper. The launch windows have equal width
and distinct endpoints, so neither contains the other. Each launch triangle
reaches the common altitude only at its own apex. The windows are
\[
W(d_1)=[-0.073352,0.323352],\qquad
W(d_2)=[-0.023352,0.373352].
\]
From the same launch time \(s=0\), the smaller-\(x\) delivery has the shorter flight duration:
\[
\ret(0,d_1)=0.193946<0.232135=\ret(0,d_2),
\]
as assumed in Case~1. Both orders are feasible, with return times
\[
(d_1,d_2):\ 0\to0.193946\to0.279361,\qquad
(d_2,d_1):\ 0\to0.232135\to0.264241.
\]
The \(x\)-ordered schedule finishes later, refuting the Case~1 claim.

\begin{figure}[!hb]
  \centering
  \includegraphics[width=\textwidth]{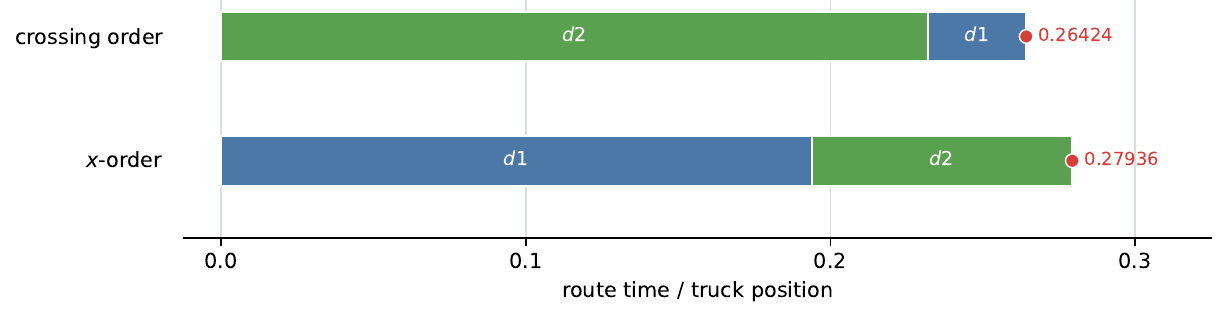}
  \caption{Case~1 counterexample. Swapping into \(x\)-order delays completion
  from \(0.264241\) to \(0.279361\).}
  \label{fig:case1-counterexample}
\end{figure}

Figure~\ref{fig:case1-counterexample} shows the two schedules.
 
\paragraph{Restricted instances with the swap property.}
The counterexamples show that properness alone does not justify restricting
attention to \(x\)-ordered schedules, so we isolate an ordering condition that suffices.
\begin{definition}[Swap-monotone instance]
An instance is \emph{swap-monotone} if, for every pair \(d_i,d_j\) with
\(x_i<x_j\) and every free time \(t\ge0\), whenever one drone can serve
\(d_j\) and then \(d_i\), it can instead serve \(d_i\) and then \(d_j\) from
\(t\), completing no later.
\end{definition}

\begin{lemma}[Ordering by adjacent swaps]\label{lem:swap-order}
Every feasible schedule on a swap-monotone instance can be transformed into
nondecreasing \(x\)-order on each drone without delaying completion.
\end{lemma}
\begin{proof}
If a drone's delivery sequence is not in nondecreasing \(x\)-order, it has an
adjacent inversion \(d_j,d_i\) with \(x_i<x_j\). Let \(t\) be the drone's free
time before this pair. Swap-monotonicity replaces \(d_j,d_i\) by \(d_i,d_j\)
from \(t\), feasibly and with no later completion. The drone can therefore
wait, if necessary, and execute the unchanged suffix at its original launch
times. This adjacent swap reduces the number of inversions by one. Repeating
the operation terminates with an \(x\)-ordered schedule and never delays its
completion.
\end{proof}

\paragraph{Dynamic programming algorithm.}
Lemma~\ref{lem:swap-order} supplies the ordering property needed by the
dynamic program.
Krizanc et al.~\cite[Thm.~4]{Krizanc2024EnRoute} define
\(T(i,j)\) using the additional condition that \(d_j\) is served last. Their
recurrence considers every possible predecessor in the \(x\)-ordered search,
resulting in \(O(n^3)\) time.
We remove this condition by incorporating the best predecessor directly into
the subproblem.

Assume that the deliveries in $D$ are indexed so that
\(x_1<\cdots<x_n\). Let \(T(i,j)\) be the earliest completion time of an
\(x\)-ordered schedule serving exactly \(i\) deliveries from
\(\{d_1,\ldots,d_j\}\). We set
$
T(0,j)=0$ 
and \(T(i,j)=+\infty\) whenever \(i>j\). We interpret
\(\ret(t,d_j)\) as \(+\infty\) if \(d_j\) cannot be served after time \(t\),
i.e., \(t>\ls(d_j)\); if \(t<\es(d_j)\), the drone waits until
\(\es(d_j)\).

The return function is nondecreasing in the drone's free time
\cite[Lem.~2]{Krizanc2024EnRoute}, so an earlier prefix completion always
dominates a later one.

 For \(1\le i\le j\le n\), the recurrence is
\[
T(i,j)
=
\min\left\{
T(i,j-1),\
\ret\bigl(\max\{T(i-1,j-1), \es(d_j)\},d_j\bigr)
\right\}.
\tag{DP}
\]
The first term skips \(d_j\), while the second serves \(d_j\) after an optimal
schedule containing \(i-1\) deliveries from
\(\{d_1,\ldots,d_{j-1}\}\).

\begin{theorem}[Quadratic-time exact algorithm]\label{thm:swap-dp}
On a swap-monotone instance with distinct \(x\)-coordinates, \textup{(DP)}
computes the exact single-drone optimum
\[
\max\{i:T(i,n)<+\infty\}
\]
in \(O(n^2)\) time.
\end{theorem}
\begin{proof}
Let \(E(i,j)\) be the earliest completion time of an \(x\)-ordered
\(i\)-delivery schedule ending at \(d_j\), or \(+\infty\) if infeasible.
Then \(T(i,j)=\min_{h\le j}E(i,h)\). For \(i\ge2\), return-time
monotonicity gives
\[
\begin{aligned}
E(i,j)&=\min_{h<j}\ret(E(i-1,h),d_j)\\
&=\ret\bigl(\min_{h<j}E(i-1,h),d_j\bigr)
=\ret(T(i-1,j-1),d_j).
\end{aligned}
\]
The final identity also holds for \(i=1\), since \(T(0,j-1)=0\).
Splitting the prefix minimum gives
\(T(i,j)=\min\{T(i,j-1),E(i,j)\}\), which is \textup{(DP)} under the
waiting convention. There are \(O(n^2)\) states and \(O(1)\) work per state.
Lemma~\ref{lem:swap-order} makes the resulting \(x\)-ordered optimum the
unrestricted optimum on swap-monotone instances.
\end{proof}

For distinct \(x\)-coordinates, \textup{(DP)} computes the optimum among
\(x\)-ordered schedules in \(O(n^2)\) time. Swap-monotonicity guarantees that
this value is the unrestricted single-drone optimum.

\section{Problem II: Minimum Fleet Size}
\label{sec:minfleet}

In this problem, every delivery must be served, and the objective is to minimize
the required number \(k^\ast\) of drones. Equivalently, \(k^\ast\) is the
smallest \(k\in\{1,\ldots,n\}\) for which a feasible full-coverage schedule
exists. The case \(k^\ast=n\), in which assigning one delivery to each drone
is optimal, is trivial; henceforth assume \(n>k^\ast\). Since computing
\(k^\ast\) exactly is NP-hard (see
Theorem~\ref{thm:nphard}), we instead apply the single-drone greedy algorithm
repeatedly until all deliveries are served. Its exact integer residual bound
and logarithmic consequence are given below.

\begin{theorem}
\label{thm:nphard}
The minimum-fleet problem is NP-hard on general instances.
\end{theorem}
\begin{proof}
In the \textsc{3-Partition} reduction of Krizanc et
al.~\cite[Thm.~1]{Krizanc2024EnRoute}, the requested schedule length equals
the total number of constructed deliveries. Thus deciding whether one drone
serves every delivery is NP-hard. This is exactly the decision
\(k^\ast\le1\), so computing \(k^\ast\) is NP-hard. This argument does
not separately establish hardness of deciding \(k^\ast\le k\) for every fixed
\(k\ge2\).
\end{proof}
Using only the single-drone \(\tfrac12\)-approximation guarantee, each round
serves at least a \(1/(2k^\ast)\) fraction of remaining deliveries. This gives
a black-box ratio bound of \(1+2\ln(n/k^\ast)+1/k^\ast\), paralleling greedy
set cover~\cite{Chvatal1979SetCover}. Our charging argument gives the
logarithmic coefficient \(1+1/k^\ast\) and the following integer residual
bound. Define
\[
r_0=n,\qquad
r_{t+1}=\left\lfloor\frac{k^\ast r_t}{k^\ast+1}\right\rfloor,\qquad
\Gamma_{k^\ast}(n)=\min\{t:r_t=0\}.
\]
For \(n=5\) and \(k^\ast=2\), this gives
\(5\mapsto3\mapsto2\mapsto1\mapsto0\) and \(\Gamma_2(5)=4\).

\begin{theorem}[Integer greedy fleet-size bound]
\label{thm:minfleet-approx}
Let \(k_{\mathrm{greedy}}\) be the number of drones used by repeatedly running
the earliest-return greedy subroutine of
Krizanc et al.~\cite{Krizanc2024EnRoute} until every delivery is served. Then
\[
k_{\mathrm{greedy}}\le\Gamma_{k^\ast}(n)
\le\left\lceil (k^\ast+1)\ln\frac{n}{k^\ast}\right\rceil+k^\ast
\]
drones. Thus its approximation ratio is at most
\(1+(1+1/k^\ast)\ln(n/k^\ast)+1/k^\ast\), and never worse than the trivial
\(n/k^\ast\) ratio obtained by using one drone per delivery.
\end{theorem}

\begin{proof}
Let \(V_t\) be the residual set before greedy schedule \(t+1\), and let
\(S_{t+1}\) be its delivery set. Restricting an optimal full-coverage solution
to \(V_t\) gives \(\mathrm{OPT}(k^\ast\text{ on }V_t)=|V_t|\). Applying
Lemma~\ref{lem:firstpass} with \(P=V_t\) and \(k=k^\ast\) gives
\[
\lvert V_t\rvert\le (k^\ast+1)\lvert S_{t+1}\rvert,
\qquad
\lvert S_{t+1}\rvert\ge
\left\lceil\frac{\lvert V_t\rvert}{k^\ast+1}\right\rceil.
\]
Consequently,
\[
\lvert V_{t+1}\rvert
\le \lvert V_t\rvert-
\left\lceil\frac{\lvert V_t\rvert}{k^\ast+1}\right\rceil
=
\left\lfloor\frac{k^\ast\lvert V_t\rvert}{k^\ast+1}\right\rfloor.
\]
Induction gives \(\lvert V_t\rvert\le r_t\), hence
\(k_{\mathrm{greedy}}\le\Gamma_{k^\ast}(n)\). Dropping the floor gives
\(r_t\le n(k^\ast/(k^\ast+1))^t\le n e^{-t/(k^\ast+1)}\). After
\(\lceil(k^\ast+1)\ln(n/k^\ast)\rceil\) greedy schedules at most \(k^\ast\)
deliveries remain, and at most \(k^\ast\) further nonempty schedules finish them.
Dividing by \(k^\ast\) and using \(\lceil x\rceil\le x+1\) gives the ratio.
Finally, every nonempty greedy schedule serves at least one delivery, so
\(r_{t+1}<r_t\) whenever \(r_t\ge1\) and hence
\(k_{\mathrm{greedy}}\le\Gamma_{k^\ast}(n)\le n\), which is the trivial
\(n/k^\ast\) ratio.
\end{proof}

\begin{corollary}[Tightness at \(k^\ast=2\) and \(n=5\)]
\label{cor:minfleet-k2-tight}
For five deliveries with optimum fleet size two, Sequential Greedy can use
four drones, attaining approximation ratio \(2\).
\end{corollary}
\begin{proof}
We exhibit an instance attaining \(\Gamma_2(5)=4\).
Set \(R=2/5\) and \(v=8/5\), so that
\(m_v=\tfrac{R}{2v}\sqrt{v^2-1}=\sqrt{39}/40\), and take five deliveries:
\[
\begin{array}{c|ccccc}
 &d_1&d_2&d_3&d_4&d_5\\\hline
x&.10&.12&.16&.20&.29\\
y&.07&.14&.12&.09&.01
\end{array}
\]
Substitution in the launch-window and return-time formulas gives the
Sequential Greedy decisions below.
\[
\begin{array}{c|c|c|c}
\text{schedule}&\text{delivery}&\text{launch}&\text{return}\\\hline
S_1&d_1&0&.122185756\\
 &d_5&.122185756&.251884271\\\hline
S_2&d_4&0&.193471019\\\hline
S_3&d_3&0&.205128205\\\hline
S_4&d_2&0&.224391567
\end{array}
\]
Thus the four greedy schedules serve exactly \(2+1+1+1=5\) deliveries.

In contrast, the following two feasible schedules serve all five deliveries:
\[
\begin{array}{c|c|c|c}
\text{schedule}&\text{delivery}&\text{launch}&\text{return}\\\hline
O_1&d_2&0&.224391567\\
 &d_4&.224391567&.446938106\\\hline
O_2&d_1&0&.122185756\\
 &d_3&.122185756&.331792107\\
 &d_5&.331792107&.473519122
\end{array}
\]
Return-time monotonicity rules out both orders of \(d_2,d_3\) on one drone:
\(\ret(0,d_2)>\ls(d_3)\) and \(\ret(0,d_3)>\ls(d_2)\).
Thus \(k^\ast=2\) and the greedy ratio is \(4/2=2\).
\end{proof}

\section{Conclusion and Open Problems}
\label{sec:conclusion} 

For maximum coverage, Sequential Greedy achieves the standard
\(1-(1-\alpha/k)^k\) guarantee with an \(\alpha\)-approximate single-drone
subroutine. The charging argument strengthens it to
\(1-(k/(k+1))^k\), approaching \(1-e^{-1}\). For \(k=2\), our analysis is tight, with a matching \(5/9\) instance. For single-drone coverage, our
counterexamples refute the claimed exchange property on proper instances.
The \(O(n^2)\)-time dynamic program
finds the unrestricted single-drone optimum on swap-monotone instances with
distinct \(x\)-coordinates.
For minimum fleet size, Sequential Greedy satisfies the integer residual bound
and gives an \(O(1+\log(n/k^\ast))\)-approximation.

Several questions remain open, including establishing matching lower bounds for
maximum coverage when \(k\ge3\), characterizing natural geometric classes of
instances that satisfy swap-monotonicity, and determining whether the
logarithmic approximation guarantee in Theorem~\ref{thm:minfleet-approx} is asymptotically tight.
Checks: \url{https://github.com/JerryMathew07/Truck_Drone_Paper}.

\appendix
\section{Counterexamples for Case~2}
\label{app:case2-counterexamples}
\begin{figure}[!htbp]
  \centering
  \includegraphics[width=\textwidth]{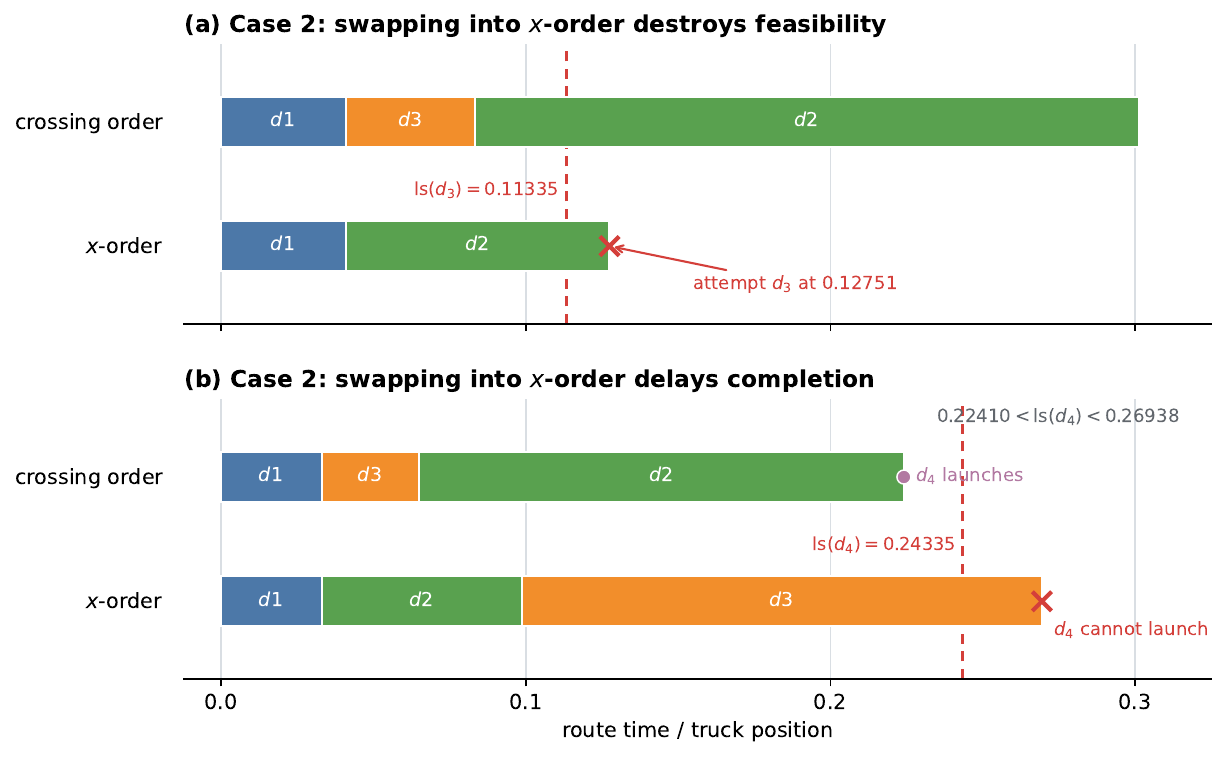}
  \caption{Case~2 counterexamples on proper instances. Swapping into
  \(x\)-order (a) makes the pair infeasible or (b) delays completion past the
  latest launch of \(d_4\).}
  \label{fig:case2-counterexamples}
\end{figure}

Both examples use \(R=0.4\), \(v=1.6\), and altitude \(y=0.02\).
Properness follows from the equal-altitude argument in
Section~\ref{sec:restricted-instances}.
Figure~\ref{fig:case2-counterexamples} shows the schedules.

\paragraph{Infeasible swap.}
Take three equally spaced deliveries with
\[
x=(0,\;0.02,\;0.04).
\]
\begin{center}
\small
\renewcommand{\arraystretch}{1.1}
\begin{tabular}{ccc}
\toprule
delivery & \(x\) & feasibility launch window \([\es,\ls]\) \\
\midrule
\(d_1\) & \(0.00\) & \([-0.32335,\ 0.07335]\) \\
\(d_2\) & \(0.02\) & \([-0.30335,\ 0.09335]\) \\
\(d_3\) & \(0.04\) & \([-0.28335,\ 0.11335]\) \\
\bottomrule
\end{tabular}
\end{center}
Serving \((d_1,d_2,d_3)\) in \(x\)-order gives
\[
d_1:\ 0.00000\to0.04103,\qquad
d_2:\ 0.04103\to0.12751,\qquad
d_3:\ \text{launch }0.12751>\ls(d_3)=0.11335.
\]
A single drone serves all three in the order \((d_1,d_3,d_2)\) with the
following times.
\[
0.00000\to0.04103\to0.08342\to0.30114.
\]
After \(d_1\), the smaller-\(x\) delivery \(d_2\) has a longer flight duration than \(d_3\)
from the same launch time. This refutes the feasibility
claim in Case~2 of Lemma~5~\cite{Krizanc2024EnRoute}.

\paragraph{Delayed completion.}
Take four deliveries with the following \(x\)-coordinates and launch windows.
\begin{center}
\small
\renewcommand{\arraystretch}{1.1}
\begin{tabular}{ccc}
\toprule
delivery & \(x\) & feasibility launch window \([\es,\ls]\) \\
\midrule
\(d_1\) & \(0.01\) & \([-0.31335,\ 0.08335]\) \\
\(d_2\) & \(0.02\) & \([-0.30335,\ 0.09335]\) \\
\(d_3\) & \(0.05\) & \([-0.27335,\ 0.12335]\) \\
\(d_4\) & \(0.17\) & \([-0.15335,\ 0.24335]\) \\
\bottomrule
\end{tabular}
\end{center}
Serve \(d_1\) first, from time \(0\), returning at \(0.033048\).
Both orders of \(d_2,d_3\) are feasible from there, with return times
\[
(d_2,d_3):\ 0.033048\to0.098759\to0.269376,
\qquad
(d_3,d_2):\ 0.033048\to0.065094\to0.224098.
\]
From the same launch time, the smaller-\(x\) delivery has the longer flight
duration, as assumed in Case~2. Swapping into \(x\)-order delays
completion past the latest launch of \(d_4\), since
\[
0.224098<\ls(d_4)=0.243352<0.269376.
\]
The inverted order serves all four deliveries and finishes at \(0.411765\).
The \(x\)-order restriction permits only three deliveries.

\end{document}